\documentclass[a4paper]{article}

\usepackage{INTERSPEECH2020, subcaption}

\title{Lite Audio-Visual Speech Enhancement}
\name{Shang-Yi Chuang$^1$, Yu Tsao$^1$, Chen-Chou Lo$^2$, Hsin-Min Wang$^3$}
\address{
  $^1$Research Center for Information Technology Innovation, Academia Sinica, Taiwan\\
  $^2$EAVISE, Dept. of Electrical Engineering, KU Leuven, Belgium \\
  $^3$Institute of Information Science, Academia Sinica, Taiwan
}
\email{\{sychuang,yu.tsao\}@citi.sinica.edu.tw, chenchou.lo@kuleuven.be, whm@iis.sinica.edu.tw}

\begin{document}

\maketitle
 
\begin{abstract}
	Previous studies have confirmed the effectiveness of incorporating visual information into speech enhancement (SE) systems. Despite improved denoising performance, two problems may be encountered when implementing an audio-visual SE (AVSE) system: (1) additional processing costs are incurred to incorporate visual input and (2) the use of face or lip images may cause privacy problems. In this study, we propose a Lite AVSE (LAVSE) system to address these problems. The system includes two visual data compression techniques and removes the visual feature extraction network from the training model, yielding better online computation efficiency. Our experimental results indicate that the proposed LAVSE system can provide notably better performance than an audio-only SE system with a similar number of model parameters. In addition, the experimental results confirm the effectiveness of the two techniques for visual data compression.
\end{abstract}

\section{Introduction}

Speech enhancement (SE) has been widely used in the front-end processing of numerous speech-related applications such as automatic speech recognition (ASR), assistive hearing technologies, and speaker recognition. The goal of SE is to improve speech quality and intelligibility by converting low-quality speech into high-quality speech. Traditional SE approaches are derived based on the assumed audio characteristics of clean speech signals and distortion sources and can be broadly divided into three categories. The first category aims to estimate the gain function to suppress noise components in noisy speech \cite{ephraim1984speech, scalart1996speech, chen2008fundamentals}. The second category aims to divide the noise and clean speech components into distinct subspaces, and the elements in the clean speech subspace are then used to generate enhanced speech \cite{rezayee2001adaptive, hu2002subspace, huang2012singing}. The third category utilizes a nonlinear model to directly map noisy speech signals to clean ones \cite{yoma1996lateral, cocchi2002subband, hussain2007nonlinear}.

Recently, deep-learning (DL) models have become popular in several fields, such as image recognition \cite{he2016deep}, ASR \cite{deng2013recent}, and natural language processing \cite{mikolov2013distributed}. In the SE field, DL models have been extensively used as fundamental units as well. Various network architectures, including the deep denoising autoencoder \cite{lu2013speech, xia2014wiener}, fully connected network \cite{liu2014experiments, xu2015regression, kolbk2017speech}, convolutional neural network (CNN) \cite{qian2017speech, fu2018end}, recurrent neural network (RNN), and long short-term memory (LSTM) \cite{weninger2014single, erdogan2015phase, sun2017multiple}, have been confirmed to notably enhance SE capabilities over those of traditional SE approaches. The success of these DL models can be attributed to their outstanding nonlinear mapping properties, which can accurately characterize complex transformations from noisy to clean speech signals.

Another advantage of DL models is their ability to fuse multimodal data. In research on SE, visual information has been adopted as auxiliary information to facilitate better SE performance \cite{hou2018audio, michelsanti2019deep, iuzzolino2020av, wang2020robust}. Compared to audio-only SE systems, audio-visual SE (AVSE) systems have been proven to provide improved enhancement capabilities. In exchange for the improved performance, however, two problems may be encountered when implementing an AVSE system. The first concerns the increase in input data. Compared to audio-only systems, the implementation of AVSE systems requires additional image inputs, which can be immense. Accordingly, additional hardware and computation costs are incurred during the training process. Second, the incorporation of face or lip images, which are particularly sensitive and personal, can generate privacy problems. Therefore, when implementing AVSE systems, particularly on embedded systems, it is essential to effectively reduce the size of visual input and user identifiability.

In this study, we propose an autoencoder (AE)-based compression network and a data compression scheme \cite{wu2018training, hsu2018study} to address these two problems in AVSE. The AE-based compression network can significantly reduce the size of the visual input while extracting highly informative visual information. A data compression scheme is applied to further reduce the bits of the extracted representation. Because we focus on reducing both the data size and training network, we refer to the proposed method as the Lite AVSE (LAVSE) system\footnote{https://github.com/kagaminccino/LAVSE}. We tested the proposed LAVSE system on the dataset of Taiwan Mandarin speech with video (TMSV)\footnote{https://bio-asplab.citi.sinica.edu.tw/Opensource.html\#TMSV}, an audio-visual version of Taiwan Mandarin hearing in noise test \cite{huang2005development}. The dataset was recorded by 18 speakers (13 males and 5 females), each providing 320 video clips. The experimental results indicate that the proposed LAVSE system can yield better performance than an audio-only SE baseline. Moreover, it is confirmed that the user identity can be removed from the compressed visual data, thereby addressing the privacy problem.

The remainder of this paper is organized as follows. Related works are described in Section \ref{sec:related_works}. The overall architecture of the proposed LAVSE system is introduced in Section \ref{sec:proposed}. The experimental setup and results are presented in Section \ref{sec:experiment}. Finally, the concluding remarks are provided in Section \ref{sec:conclusion}.

\section{Related Works}
\label{sec:related_works}

In this section, previous research works related to the proposed LAVSE system are reviewed.

\subsection{Audio-only SE}

DL-based SE systems can be broadly divided into two categories: masking- and mapping-based. For both categories \cite{wang2018supervised}, paired audio data of clean and noisy speech utterances are prepared in the training phase. The noisy speech utterances are used as the input, and the goal is to obtain enhanced speech signals as the output. During training, a loss function is used, such as L1 and L2 norms, to measure the difference between the enhanced and clean signals. The DL-based model is trained with the aim of minimizing the loss function. Although the goals are the same, the implementations of the masking- and mapping-based SE approaches are different. The masking-based approaches estimate a mask and then multiply it with the noisy audio signal to generate an enhanced signal \cite{williamson2015complex}. In contrast, mapping-based approaches directly estimate enhanced speech. According to the type of input, the methods are based on spectral mapping \cite{liu2014experiments, xu2015regression, kolbk2017speech, lu2013speech}, complex spectral mapping \cite{fu2017complex, strake2020fully}, and waveform mapping \cite{fu2018end, pandey2019new}.

\subsection{AVSE}

\begin{figure}[t]
	\centering
	\includegraphics[width=0.77\linewidth]{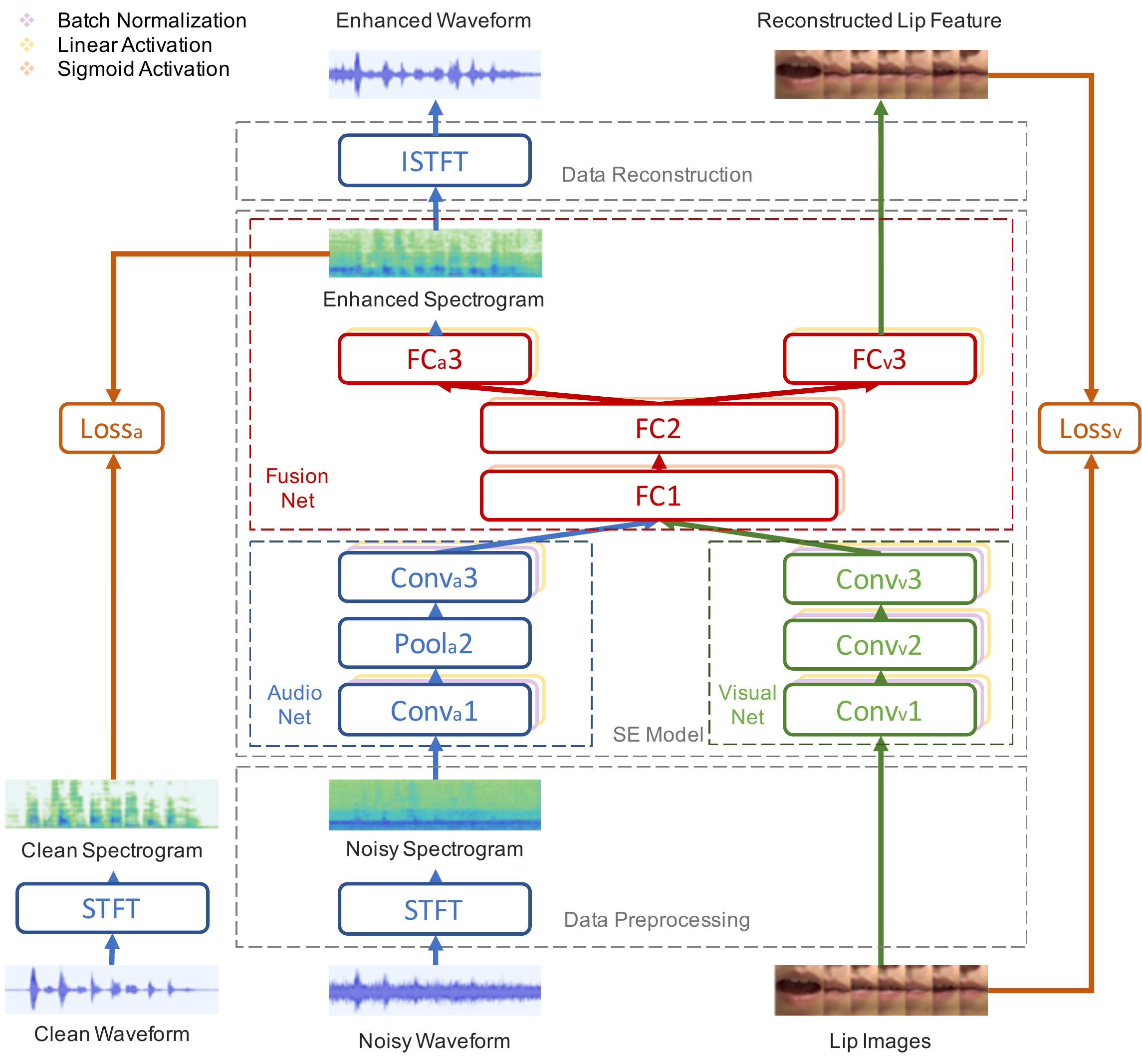}
	\caption{The AVDCNN architecture\cite{hou2018audio}.}
	\label{fig:AVDCNN}
\end{figure}

The concept of AVSE is to incorporate visual input as auxiliary data into audio-only SE systems to provide complementary information. Numerous prior studies have verified the effectiveness of using visual input in improving SE performance \cite{hou2018audio, michelsanti2019deep, iuzzolino2020av, wang2020robust}. Among them, as depicted in Fig. \ref{fig:AVDCNN}, audio-visual speech enhancement using multimodal deep convolutional neural networks (AVDCNN) \cite{hou2018audio} receives noisy audio and lip images as the input and generates enhanced audio and lip images as the output. The combination of audio loss, Loss{\footnotesize a} and visual loss, Loss{\footnotesize v} (with specific weights) is used as the final loss to optimize the AVSE model. In this study, we used AVDCNN (with the audio and fusion net of the LAVSE system) as the basic AVSE system. We derived several algorithms to reduce the system’s computation cost and storage requirements and address privacy problems.

\subsection{Bit-wise Data Compression}

The single-precision floating-point in IEEE 754 \cite{institute1985ieee} is a data format widely used to represent 32-bit floating-point variables. The 32 bits consist of 1 sign bit, 8 exponential bits, and 23 mantissa bits. The sign bit indicates whether the value represented is positive or negative, the exponential bits determine the representation range of the value, and the mantissa bits account for the significant figures. The exponential and mantissa terms are not explicitly assigned and should be computed using addition and multiplication.

Because quantization on the mantissa bits does not change the represented value itself but only reduces the precision, a previous work suggests an exponent-only floating-point (EOFP) \cite{hsu2018study} format for audio SE tasks. Based on this design, the model size can be notably reduced, and the online computation efficiency can be effectively enhanced \cite{lin2019ia} while maintaining the SE performance in terms of speech quality and intelligibility.

\section{Proposed LAVSE System}
\label{sec:proposed}

The goal of this study is to reduce the amount of storage, enhance the online computation efficiency, and address privacy problems of the AVDCNN system, which is depicted in Fig. \ref{fig:AVDCNN}. In this section, we first introduce the overall architecture of the proposed LAVSE system and then detail two approaches for reducing the amount of visual data.

\subsection{Overall Architecture}

\begin{figure}[t]
	\centering
	\includegraphics[width=0.77\linewidth]{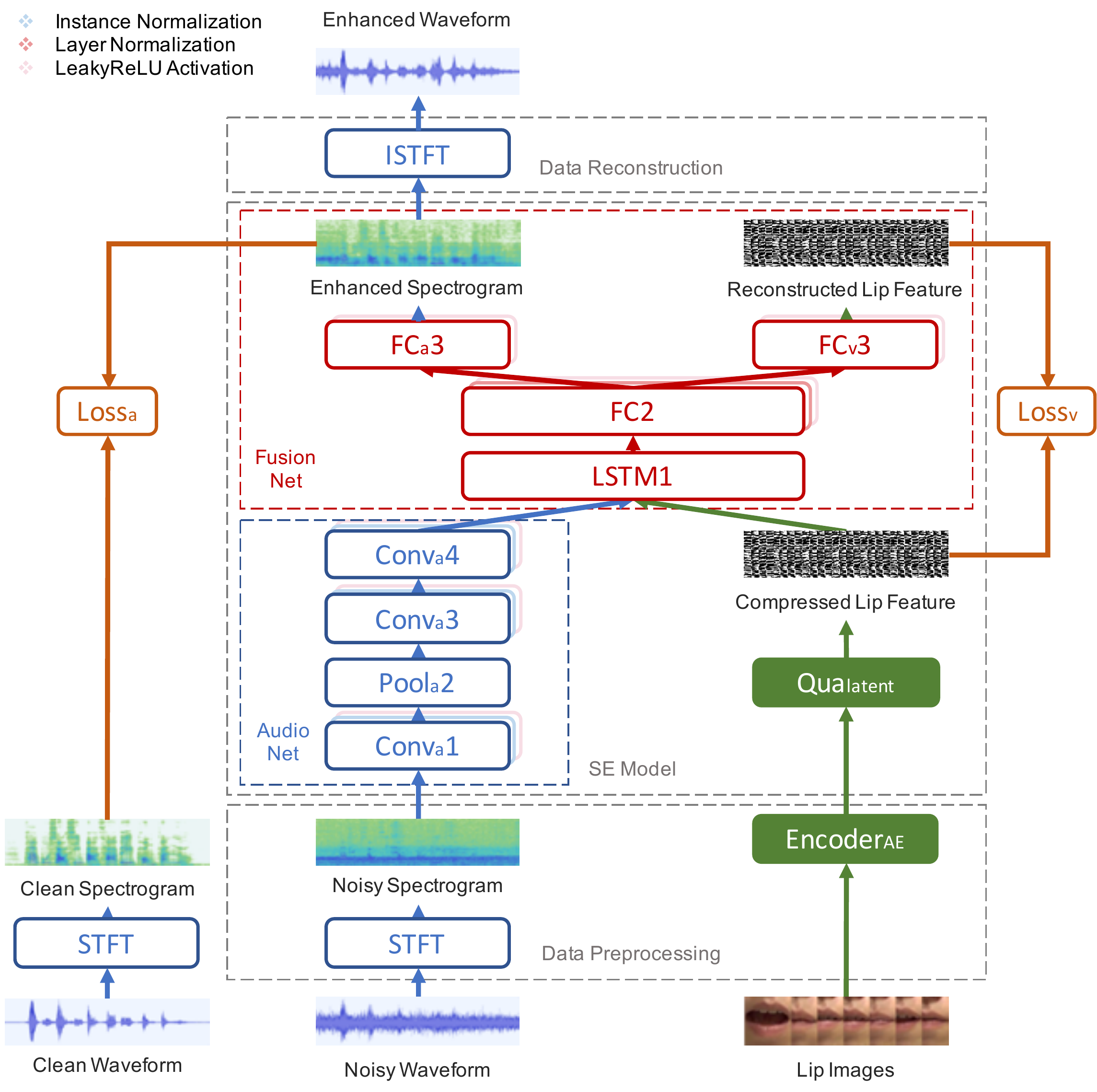}
	\caption{The LAVSE architecture with two visual data compression units (Encoder{\footnotesize AE} and Qua{\footnotesize latent}).}
	\label{fig:whole}
\end{figure}

In Fig. \ref{fig:whole}, the architecture of the LAVSE system is depicted, which includes three parts: data preprocessing, SE model, and data reconstruction. The two green solid blocks outline the proposed visual data compression. Encoder{\footnotesize AE} reduces the dimension of the visual data, whereas Qua{\footnotesize latent} conducts bit quantization.

\begin{figure}[t]
	\centering
	\includegraphics[width=0.75\linewidth]{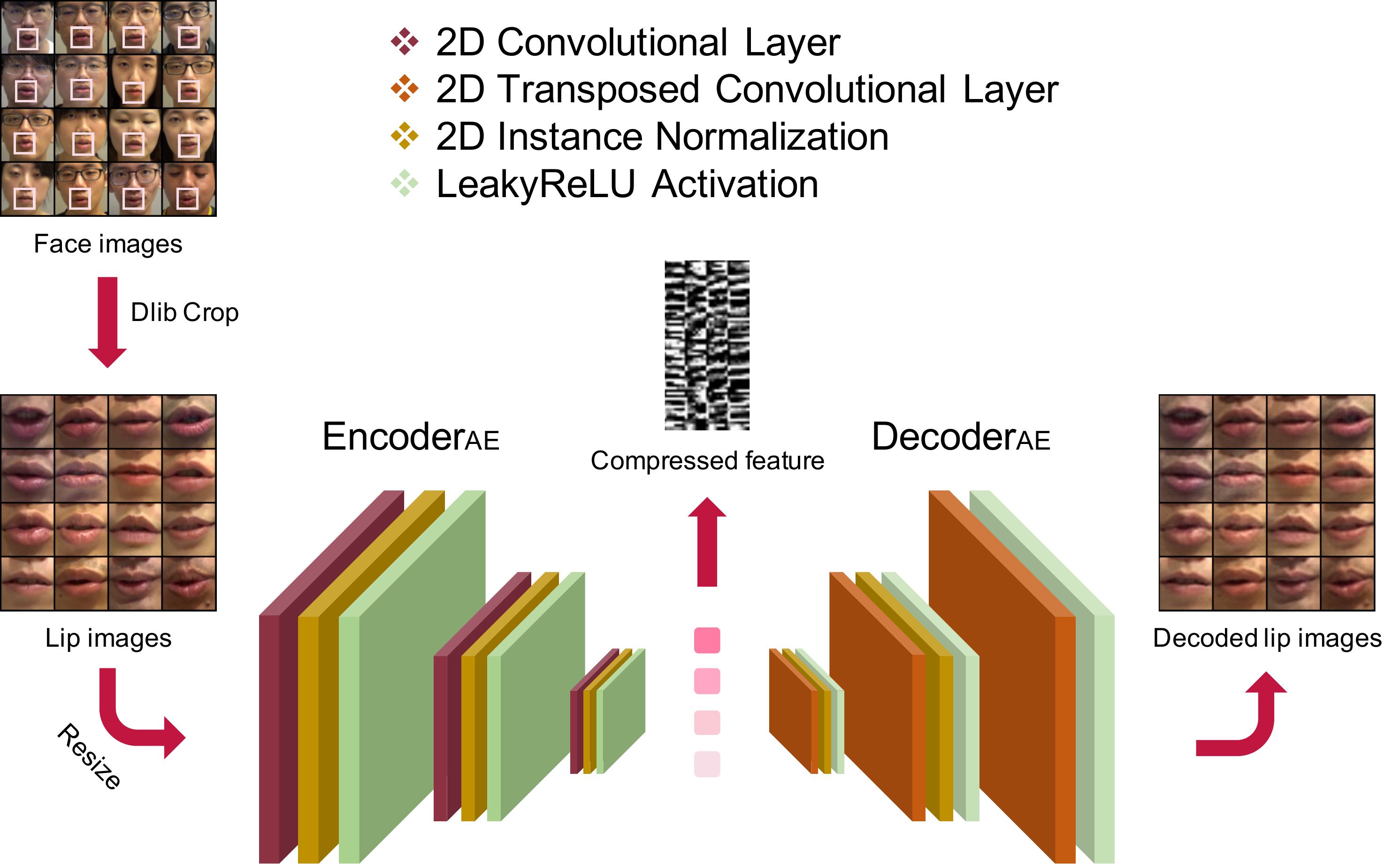}
	\caption{
	The AE model for visual data compression.}
	\label{fig:AE}
\end{figure}

Before training, noisy waveforms are transformed into audio features using the short-time Fourier transform (STFT). Lip images are fed into the AE model illustrated in Fig. \ref{fig:AE}, in the preprocessing stage, and the latent representations are used as the visual features. With Encoder{\footnotesize AE}, the visual feature extraction net can be fully removed from the training model. Next, in the SE model stage, the audio features are initially enhanced, whereas the visual features are processed by Qua{\footnotesize latent} to reduce the precision. The processed audio and visual features are then concatenated and fed into a fusion net, which is formed using LSTM and FC layers. Finally, enhanced audio features and restored visual features are obtained. With reference to clean audio data and compressed visual data, we can estimate the combined loss (from the audio and visual parts) and then use it to update the model parameters. 

During testing, noisy waveforms and lip images first go through the preprocessing stage, and then they are fed into the trained SE model to generate enhanced spectrograms, which are then converted into the waveform domain through inverse STFT with the phases of the noisy audio in the reconstruction stage. 

\subsection{Visual Data Compression}

The visual data compression module consists of two parts. The original lip images were first processed by Encoder{\footnotesize AE}. The AE model is trained in a self-learning manner in advance, which consists of three 2D convolutional layers as the encoder and three 2D transposed convolutional layers as the decoder. The dimensions of the latent representation of the AE are 32 $\times$ 8 $\times$ 8 which yield 2048 dimensions after flattened. This latent representation is used as the visual feature, and its size is only 16.67\% of that of the original lip image (3 $\times$ 64 $\times$ 64).

Next, the extracted visual features are processed by Qua{\footnotesize latent} (based on EOFP) to further reduce the number of bits for each feature element. The compressed visual features are then used in the SE model. In real-world applications, the Encoder{\footnotesize AE} and Qua{\footnotesize latent} modules can be installed in the visual sensor so that the online computation efficiency is enhanced. Through the two-stage visual data compression process, the transmission cost can be reduced significantly and the lip images can be blurred out so that the matter of privacy can be addressed moderately.

\section{Experiments}
\label{sec:experiment}

This section first introduces the experimental setup, including data preparation and system settings, and then presents the experimental results and discussions.

\subsection{Experimental Setup}

The TMSV dataset contained 5,760 speech utterances from 18 speakers, each providing 320 utterances. As there were only 5 females, we used 5 speakers from each gender to create a gender-balanced situation. The training set consisted of the 1st to the 200th utterance from each of the eight speakers (four males and four females), amounting to 1600 clean utterances in all. These utterances were corrupted by 100 types of noise \cite{hu2004100} at five signal-to-noise ratios (SNRs) from -12 dB to 12 dB with a step of 6 dB. The test set consisted of the 201st to the 320th utterance from each of the other two speakers (one male and one female) for a total of 240 utterances. We designed a car driving application scenario for which we adopted six types of noise that are more common in car driving, including the cries of a baby, engine noise, background talkers, music, pink noise, and street noise, to form the noisy testing data. The SNR levels were set to -1, -4, -7, and -10 dB, which are considered challenging. Note that the speakers, noise types, and SNR levels are all mismatched in the training and testing settings.

Two standard evaluation metrics were used to evaluate the performance: perceptual evaluation of speech quality (PESQ) \cite{rix2001perceptual} and the short-time objective intelligibility measure (STOI) \cite{taal2011algorithm}. PESQ was used to evaluate the quality of speech, with a score ranging from -0.5 to 4.5. STOI was designed to evaluate the intelligibility of speech, with a score ranging from 0 to 1. Higher PESQ and STOI scores indicate that the enhanced speech has better speech quality and intelligibility, respectively.

We adopted the log1p magnitude spectrum as the audio feature for the SE model\footnote{According to our preliminary experiments, the log1p magnitude spectrum can yield better SE performance than the log power spectrum.}. For the visual part, the lip or face image contours were positioned using Dlib \cite{dlib09}, which is a 68-point facial landmark detector, and then compressed using Encoder{\footnotesize AE}. The audio-visual combined loss is $loss = loss_{a} + \mu \times loss_{v}$, where $\mu$ is empirically determined as $10^{-3}$ and the individual losses are mean square errors. We have attempted several visual feature extraction approaches to build the baseline AVSE system, which will be detailed in Section \ref{sec:baseline}. The audio-only SE system was implemented for comparison. The total number of model parameters of the audio-only SE system were designed to be comparable to that of the AVSE system. For all the SE models, the optimizer is Adam \cite{kingma2014adam} with a learning rate of $5 \times 10^{-5}$.

\subsection{Experimental Results}

\subsubsection{LAVSE versus Audio-only SE}
\label{sec:baseline}

First, we intend to compare the performance of the proposed LAVSE with Encoder{\footnotesize AE} (LAVSE(AE)) with several baseline systems, including an audio-only SE (Audio-only) system and three AVSE systems with different visual data, namely, (1) AVSE(VGGface)---face features processed by VGGface \cite{parkhi2015deep}, (2) AVSE(face)—--raw face images, and (3) AVSE(lip)—--raw lip images. The performances of these five systems are presented in Table \ref{tab:baseline}, where the scores of the original noisy speech are also listed for comparison. The results indicate that Audio-only, AVSE(lip), and LAVSE(AE) can notably improve the speech quality (in terms of PESQ) and intelligibility (in terms of STOI). However, AVSE(VGGface) and AVSE(face) cannot yield further improvements compared to the Audio-only system. This may be because the entire face image may contain redundant information that does not directly benefit the SE task. It is also noted that LAVSE(AE) outperforms Audio-only and AVSE(lip) in terms of both PESQ and STOI, suggesting the positive effect of incorporating compressed visual lip information into the SE system. In the following discussion, we present only the results obtained using lip images as visual data.

\begin{table}[th]
	\caption{PESQ and STOI scores of the LAVSE(AE) system and the baselines.}
	\label{tab:baseline}
	\centering
	\begin{tabular}{ccc}
		\toprule
		& \textbf{PESQ} & \textbf{STOI}  \\
		\midrule
		\textbf{Noisy}         & 1.001          & 0.587          \\
		\textbf{Audio-only}    & 1.283          & 0.610          \\
		\textbf{AVSE(VGGface)} & 0.797          & 0.492          \\
		\textbf{AVSE(face)}    & 1.270          & 0.616          \\
		\textbf{AVSE(lip)}     & 1.337          & 0.641          \\
		\textbf{LAVSE(AE)}     & \textbf{1.374} & \textbf{0.646} \\
		\bottomrule
	\end{tabular}
\end{table}

\subsubsection{Visual Data Compression}

\begin{figure}[t]
	\centering
	\begin{subfigure}[t]{0.45\linewidth}
		\centering
		\includegraphics[scale=0.25]{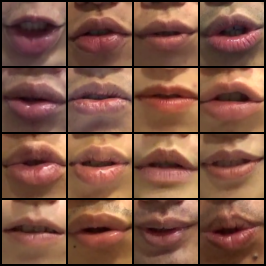}
		\caption{Original lip images.}
		\label{fig:lip_ori}
	\end{subfigure}
	\begin{subfigure}[t]{0.45\linewidth}
		\centering
		\includegraphics[scale=0.25]{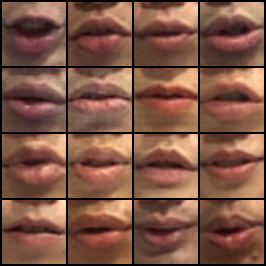}
		\caption{AE reconstructed images.}
		\label{fig:lip_com}
	\end{subfigure}
	\caption{Original and AE reconstructed lip images.}
	\label{fig:lip_img}
\end{figure}

Next, we investigate the effects of the two visual data compression modules: Encoder{\footnotesize AE} and Qua{\footnotesize latent}. The original lip images and the lip images that are put through the AE are presented in Fig. \ref{fig:lip_img}. Through visual inspection, we can affirm that the AE can encode and decode lip images well, suggesting that the latent representation can carry enough lip image information. The compression ratio of the Encoder{\footnotesize AE} module $R_{\text{AE}}$ was 6 ($= (3 \times 64 \times 64) \div 2048$).

\begin{figure}[t]
	\centering
	\includegraphics[width=\linewidth]{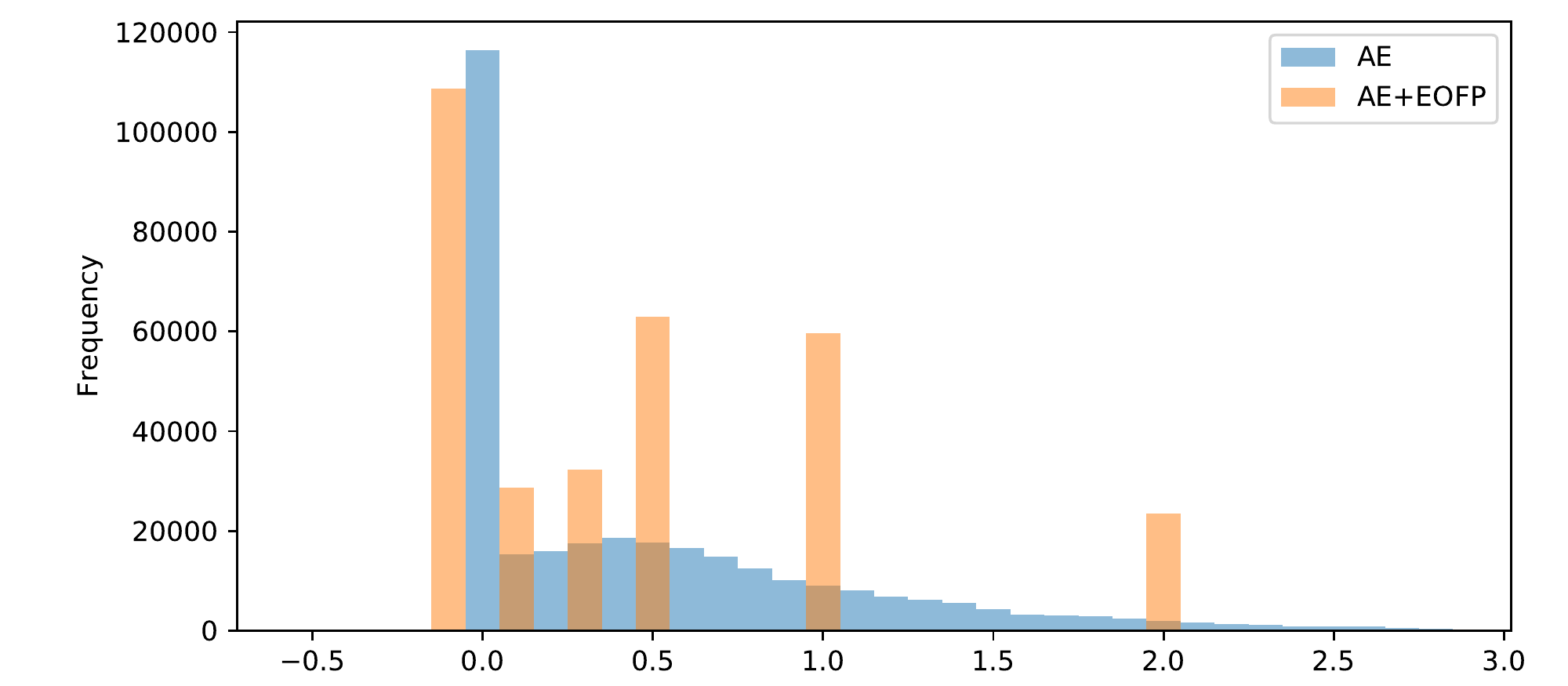}
	\caption{The distributions of visual features before and after applying Qua{\footnotesize latent}.}
	\label{fig:latent_freq}
\end{figure}

\begin{figure}[t]
	\centering
	\begin{subfigure}[t]{0.45\linewidth}
		\centering
		\includegraphics[scale=1.1]{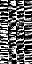}
		\caption{AE feature.}
		\label{fig:latent_full}
	\end{subfigure}
	\begin{subfigure}[t]{0.45\linewidth}
		\centering
		\includegraphics[scale=1.1]{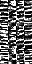}
		\caption{AE+EOFP feature.}
		\label{fig:latent_eofp}
	\end{subfigure}
	\caption{Visual latent features of lips.}
	\label{fig:latent}
\end{figure}

Subsequently, we investigate the effect of Qua{\footnotesize latent} by comparing the visual features before and after applying Qua{\footnotesize latent}. In this study, we applied Qua{\footnotesize latent} to reduce the original 32-bit data point to a 4-bit representation, including 1 sign bit and 3 exponential bits. The compressed representation was then used as the visual feature and put through the SE model. Note that by applying Qua{\footnotesize latent} to the latent representation, the sizes of the visual features were notably reduced—--each feature element was reduced from 32 to 4 bits. The compression ratio of the Qua{\footnotesize latent} module $R_{\text{Qua}}$ is 8 ($= (1+8+23) \div (1+3+0)$). The LAVSE system with both Encoder{\footnotesize AE} and Qua{\footnotesize latent} is termed LAVSE(AE+EOFP). By comparing the distributions presented in Fig. \ref{fig:latent_freq}, we observe that when using Qua{\footnotesize latent}, the distribution of the compressed visual features (orange bars) covers the entire distribution of uncompressed features (blue bars) well. In Figs. \ref{fig:latent_full} and \ref{fig:latent_eofp}, the latent representations of lip features before and after applying Qua{\footnotesize latent}, respectively, are depicted. From the figures, we can easily see that the user identity has been removed almost completely, thereby moderately addressing the privacy problem.

\begin{figure}[t]
	\begin{subfigure}[t]{\linewidth}
		\centering
		\includegraphics[width=0.9\linewidth]{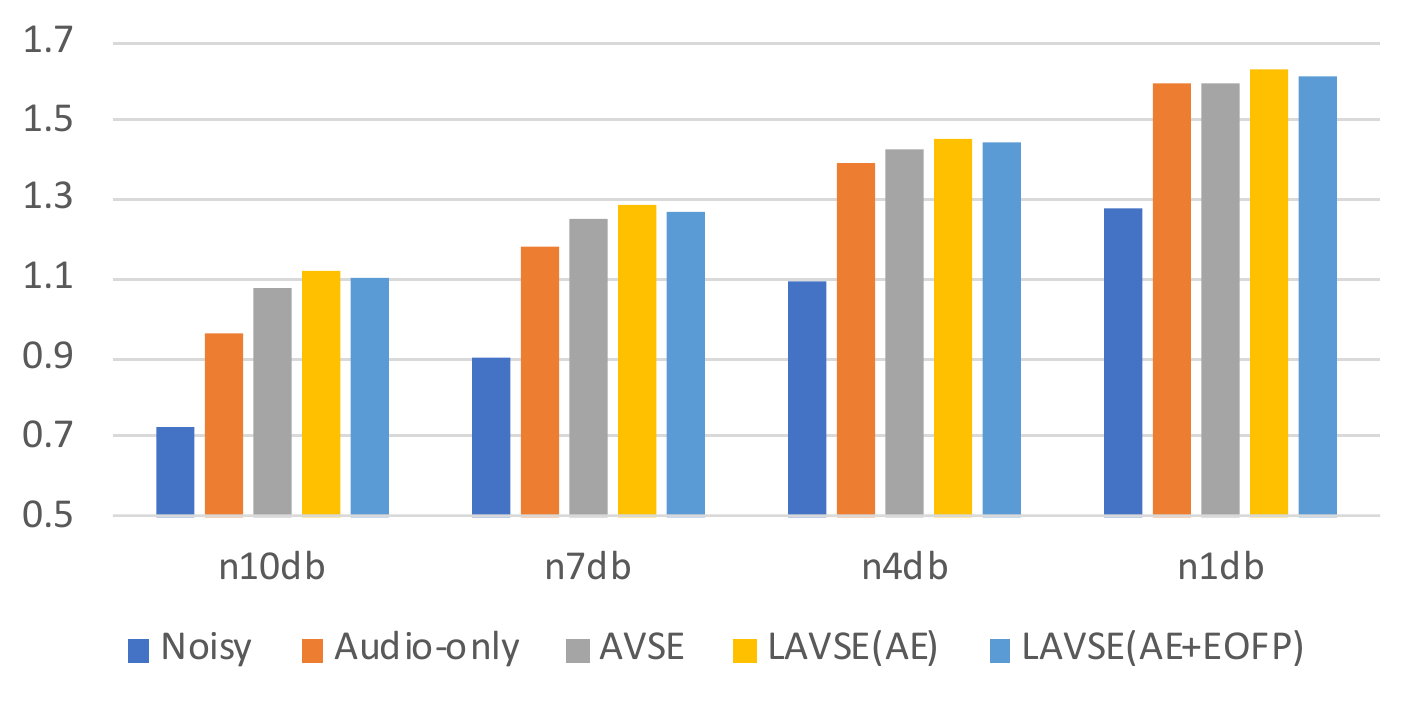}
		\caption{PESQ.}
	\end{subfigure}
	\centering
	\begin{subfigure}[t]{\linewidth}
		\centering
		\includegraphics[width=0.9\linewidth]{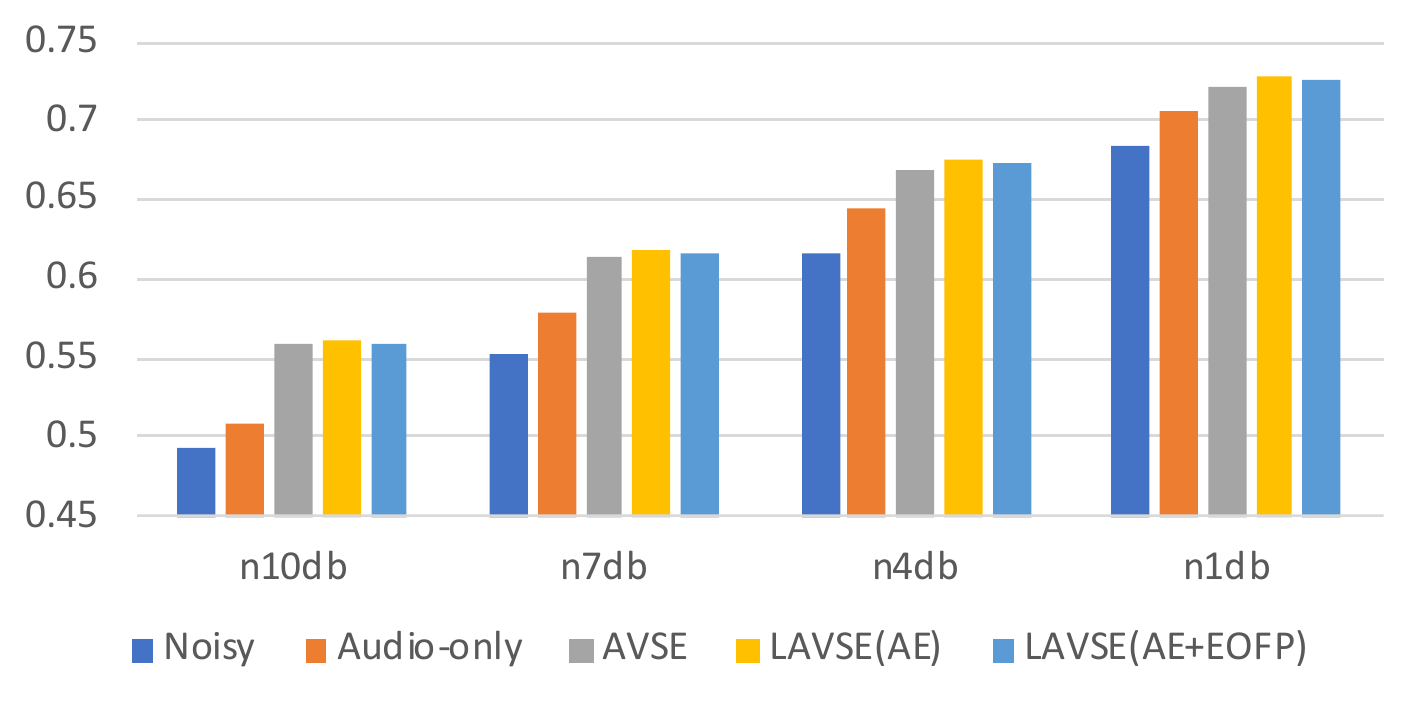}
		\caption{STOI.}
	\end{subfigure}
	\caption{PESQ and STOI scores at specific SNR levels.}
	\label{fig:performance}
\end{figure}

Finally, by combining Encoder{\footnotesize AE} and Qua{\footnotesize latent}, the overall compression ratio, $R_{\text{Comp}}$ is 48 ($= R_{\text{AE}} \times R_{\text{Qua}}$), confirming that the size of the visual input can be significantly reduced using the LAVSE(AE+EOFP) system. From Fig. \ref{fig:performance}, it is evident that even with such significant data compression, LAVSE(AE+EOFP) can still yield satisfactory enhancement performance of PESQ = 1.358 and STOI = 0.643, which are comparable to those of the best AVSE systems and notably better than those of the Audio-only system (refer to Table \ref{tab:baseline}).

\section{Conclusion}
\label{sec:conclusion}

In this study, a LAVSE system with novel visual data compression approaches was proposed to address the problems that may be encountered when implementing an AVSE system for practical applications. We designed a lighter model structure by removing the visual feature extraction network from the training model. Compared to the data sizes of the original lip images, the data sizes of the visual features obtained using the combined Encoder{\footnotesize AE} and Qua{\footnotesize latent} modules are reduced by a significant factor of 48. Experimental results confirm that the AVSE system with the proposed visual data compression provides better performance than the Audio-only and AVSE systems without using visual data compression. The contributions of this study are threefold. First, we verified the effectiveness of incorporating visual information into SE. Second, we confirmed that even if the visual data is significantly reduced, it can still provide significant complementary information for the SE task. Third, using the Encoder{\footnotesize AE} and Qua{\footnotesize latent} modules together, privacy problems may be addressed as well. We expect that the findings in this study would be helpful for implementing the DL-based AVSE model in real-world applications. In the future, we will investigate time-domain compression to further enhance the online computation efficiency of the proposed LAVSE system.

\bibliographystyle{IEEEtran}
\bibliography{refs}

\end{document}